\documentclass[prl, twocolumn, nofootinbib, amsmath, amssymb]{revtex4}
\usepackage{graphics,hyperref}

\begin{document}
\date{\today}
\title{Completely positive maps and classical correlations}
\author{C\'{e}sar A. Rodr\'{i}guez-Rosario} 
\email[email: ]{carod@physics.utexas.edu}
\affiliation{  The University of Texas at Austin, Center for Complex Quantum Systems, 1 University Station C1602, Austin TX 78712}
\author{Kavan Modi} 
\affiliation{  The University of Texas at Austin, Center for Complex Quantum Systems, 1 University Station C1602, Austin TX 78712}
\author{Aik-meng Kuah} 
\affiliation{  The University of Texas at Austin, Center for Complex Quantum Systems, 1 University Station C1602, Austin TX 78712}
\author{Anil Shaji}
\affiliation{ Department of Physics and Astronomy, University of New Mexico,
Albuquerque NM 87131}
\author{E.~C.~G. Sudarshan}
\affiliation{  The University of Texas at Austin, Center for Complex Quantum Systems, 1 University Station C1602, Austin TX 78712}

\begin{abstract}

We expand the set of initial states of a system and its environment that are
known to guarantee completely positive reduced dynamics for the system when the
combined state evolves unitarily. We characterize the correlations in the
initial state in terms of its quantum discord [H.~Ollivier and W.~H.~Zurek,
Phys.~Rev.~Lett.~{\bf 88}, 017901 (2001)]. We prove that initial
states that have only classical correlations lead to completely positive reduced
dynamics. The induced maps can be not completely positive when quantum
correlations including, but not limited to, entanglement are present. We
outline the implications of our results to quantum process tomography
experiments. 

\end{abstract}


\maketitle 

In the mathematical theory of open quantum systems
\cite{davies76a,spohn80a,breuer02a} it is often assumed that the system of
interest and its environment are initially in a product state. This extremely
restrictive assumption precludes the theory from describing a wide variety of
experimental situations including the one in which an open system is simply
observed for some interval of time without attempting to initialize it in any
particular state at the beginning of the observation period. If dynamical maps
\cite{PhysRev.121.920} are used to describe the open evolution, then it is known
that an initial product state leads to dynamics of the system 
described in terms of completely positive maps
\cite{choi72a,choi75,kraus83b,sudarshan86,SudChaos}. There has been significant
experimental and theoretical interest in quantum correlations, entanglement and
coherence in the context of quantum information theory \cite{Nielsen00a}. It is
only recently that interest has picked up in investigating how these properties,
when present in the initial state of a system and its environment, affects the
open evolution of the system
\cite{pechukas94a,jordan05a,shaji05dis,jordan:052110,terno05,shabani06a,ziman06}. In this Letter we investigate the related question of how to relax the initial
product state assumption and still obtain dynamics for the system that are
described by completely positive transformations. 

Consider a generic bipartite state $\rho^{\mathcal{SE}}$ of a quantum system
$\mathcal{S}$ and its environment $\mathcal{E}$. Unitary evolution of
the combined state induces transformations on the system state described
by a dynamical map $\mathfrak{B}$ defined as
\begin{equation}\label{TotalTrace} 
\eta\rightarrow\mathfrak{B}(\eta)\equiv\mbox{Tr}_{\mathcal{E}}\left[U
\rho^{\mathcal{SE}} U^\dag\right]=\eta^{\prime},
\end{equation}
where $\eta = {\mbox{Tr}}_{\mathcal{E}} \rho^{\mathcal{SE}}$ is the
initial state of ${\mathcal{S}}$ and $\eta'$ is its final state. We use $\eta$
to represent density matrices of the system $\mathcal{S}$ and $\tau$ to
represent density matrices of the environment $\mathcal{E}$. The action of the
map can be written in terms of its eigenmatrices $\{\zeta^{( \alpha )} \}$ and
eigenvalues $\{\lambda_\alpha\}$,
\begin{equation*}\label{oper1}
\mathfrak{B}(\eta)=\sum_{\alpha}\lambda_{\alpha}\;\zeta^{(\alpha)}\eta
\;{\zeta^{(\alpha)}}^{\dagger}.\\
\end{equation*}

If the initial state of the system and its environment is simply separable
(product) so that $\rho^{\mathcal{SE}}=\eta \otimes \tau$, then the
eigenvalues of the dynamical map are all positive for any choice of unitary
evolution \cite{sudarshan86,SudChaos}. In this case we can define
$C^{(\alpha)}\equiv\sqrt{\lambda_{\alpha}}\zeta^{(\alpha)}$ to get
\begin{equation}\label{oper2}
\mathfrak{B}(\eta)=\sum_{\alpha}C^{(\alpha)}\eta {C^{(\alpha)}}^{\dagger},
\end{equation}
with $\sum_{\alpha}{C^{(\alpha)}}^{\dagger}C^{(\alpha)}=1$. Any map that can be
written in this form is completely positive \cite{choi72a,choi75}.  
If the initial system and environment state is not a product state,
then the map induced by arbitrary unitary evolution on $\rho^{\mathcal{SE}}$ is
in general not completely positive \cite{pechukas94a,jordan05a}. In
this Letter we identify a general class of initial states such that 
\emph{any} unitary transformation on it leads to completely positive reduced
dynamics for the system. Simply separable states are a subset of this general
class of states. To characterize this class we use the notion of
quantum discord introduced by Ollivier and Zurek \cite{PhysRevLett.88.017901}
and independently by Henderson and Vedral \cite{henderson01a}. 

First, we consider an example that shows how not completely positive dynamics 
arise in physically realizable situations where the initial state is separable
but not simply separable. Let ${\mathcal{S}}$ and ${\mathcal{E}}$ 
both be qubits in a combined initial state,  
\begin{equation}\label{sep1}
\rho^{\mathcal{SE}} =  \frac{1}{4} (\openone\otimes\openone+a_j
\sigma_j\otimes\openone+c_{23}\sigma_2\otimes\sigma_3),
\end{equation}
where $j =1,2,3$, $\sigma_j$ are the Pauli matrices and repeated
indices are summed over. To show that $\rho^{\mathcal{SE}}$ is a separable state
we use the Peres partial transpose test \cite{peres96a} which is a necessary and
sufficient test for entanglement in two qubit systems. The transpose operation
takes $\sigma_2$ to $ -\sigma_2$ while leaving the other two Pauli matrices
intact. If we apply the partial transpose test to $\rho^{\mathcal{SE}}$ by
transposing ${\mathcal{E}}$, we see from Eq.~(\ref{sep1}) that
$(\rho^{\mathcal{SE}})^{\mbox{PT}} = \rho^{\mathcal{SE}}$ and so it is a
separable state. The initial state of the system
is $\eta=\mbox{Tr}_\mathcal{E}[\rho^{\mathcal{SE}}] =
( \openone+a_j \sigma_j )/2$. 

Consider a unitary evolution of
$\rho^{\mathcal{SE}}$ given by  $U = e^{-iHt} = \cos(\omega
t)\openone\otimes\openone - i \sin(\omega t) \sigma_j
\otimes \sigma_j$, where $H = \omega \sum_j \sigma_j \otimes \sigma_j$. The
state of the system at time $t$ is given by \cite{rodr,jordan06a}
\begin{eqnarray*}
\eta' 
&=& \frac{1}{2}\big[ \openone + \cos^2\left(2\omega t\right) a_j
\sigma_j  + c_{23}\cos\left(2\omega t\right) \sin\left(2\omega
t\right)\sigma_1 \big].
\end{eqnarray*}

The dynamical map $\mathfrak{B}$ that describes the open evolution of the
system qubit ${\mathcal{S}}$ is an affine transformation \cite{jordan:034101}
that squeezes the Bloch sphere of the qubit into a sphere of radius $\cos^2 (2
\omega t)$ and shifts its center by $c_{23}\cos(2\omega t)\sin(2 \omega t)$ in
the $\sigma_1$ direction. The eigenvalues of the map are
\begin{eqnarray*} 
\lambda_{1,2}&=&\frac{1}{2}\big[1-\cos^2 (2\omega t)\pm c_{23}\cos (2 \omega
t)\sin (2 \omega t ) \big], \\
\lambda_{3,4}&=&\frac{1}{2}\Big[1+ \cos^2 (2\omega t ) \nonumber
\\
&& \hspace{5 mm} \pm \cos (2\omega t )\sqrt{ 4\cos^2 ( 2\omega
t )+c_{23}^{2} \sin^2 ( 2\omega t )} \Big].
\end{eqnarray*}
It is easily seen that $\lambda_{3,4}$ are always positive, while for
$\lambda_{1,2}$ to be positive we need $\sin^2 (2\omega t ) \geq \pm c_{23}\cos
(2\omega t)\sin (2\omega t)$. We can choose $c_{23}$ such that this condition
will be violated for some values of $\omega t$ making the map $\mathfrak{B}$ not
completely positive. It has been previously shown that not completely positive
maps come from initial entanglement \cite{jordan:052110}. This example shows
that even separable states can lead to not completely positive maps. A similar
example has been worked out in \cite{terno05}. The map $\mathfrak{B}$ has a
physical interpretation as long as it is applied to initial states $\eta$ that
are {\em compatible} with the total state $\rho^{\mathcal{SE}}$
\cite{shaji05dis}.  

From this example we see that correlations, and not necessarily entanglement, in
the initial state of the system and its environment can lead to not completely
positive reduced dynamics for ${\mathcal{S}}$. Do all correlations lead to not
completely positive maps? If not, is there a way of characterizing these
correlations that lets us easily see if a given initial state will lead to
completely positive dynamics or not? 

The traditional division of bipartite density matrices $\rho^{\mathsf{XY}}$
into separable ($\rho^{\mathsf{XY}}=\sum_j p_j \eta_j \otimes \tau_j $) and
entangled is often taken to be synonymous with classical correlations and
quantum correlations respectively \cite{PhysRevA.40.4277}. Ollivier and Zurek
\cite{PhysRevLett.88.017901} and independently Henderson and Vedral
\cite{henderson01a} have proposed a different definition for classical and
quantum correlations in density matrices based on information theoretic
considerations. Suggestions for characterizing the correlations along similar
lines were also made by Bennett et al. in \cite{bennett99b,bennett02a}.

To quantify the correlations between two systems $\mathsf{X}$ and $\mathsf{Y}$,
we can either compute the mutual information,
\begin{equation}
\label{mutual}
\mathbf{I}(\mathsf{Y}:\mathsf{X})=\mathbf{H}(\mathsf{X})+\mathbf{H}(\mathsf{Y}
)-\mathbf{H}(\mathsf{X}\cup\mathsf{Y}),
\end{equation}
or its classical equivalent,
\begin{equation}
\label{conditional}
\mathbf{J}(\mathsf{Y}:\mathsf{X})=\mathbf{H}(\mathsf{Y})-\mathbf{H}(\mathsf{Y}
|\mathsf{X}),
\end{equation}where $\mathbf{H}$ is the Shannon
entropy \cite{shannon48a}. 
If $\mathsf{X}$ and $\mathsf{Y}$ are classical systems in the
sense that their states are described by probability distributions over two 
random variables $\mathsf{X}$ and $\mathsf{Y}$, then $\mathbf{J}=\mathbf{I}$ as a
consequence of Bayes' rule.  

If $\mathsf{X}$ and $\mathsf{Y}$ are quantum systems with their state described
by the density matrix $\rho^{\mathsf{XY}}$, then the mutual information between
the two can be computed by replacing the Shannon entropy in Eq.~(\ref{mutual})
by the von Neumann entropy ${\mathbf H}(\rho)= -{\mbox{Tr}} \rho \log \rho$ \cite{vonNeu}. To
compute $\mathbf{J}(\mathsf{Y}:\mathsf{X})$, the definition of
$\mathbf{H}(\mathsf{Y}|\mathsf{X})$ has to be generalized to
\begin{equation}
\label{gen1}
{\mathbf H}\big({\mathsf Y} \big| \big\{ \Pi^{\mathsf{X}}_j \big\} \big) = \sum
_j p_j {\mathbf H} ( \rho_{{\mathsf Y}| \Pi^{\mathsf X}_j} ), 
\end{equation}
where $p_j = {\mbox{Tr}}_{\mathsf{X}, \, \mathsf{Y}} \Pi^{\mathsf X}_j
\rho^{\mathsf{XY}}$ and $\rho_{\mathsf{Y}|\Pi^{\mathsf Y}_j}= 
\Pi^\mathsf{X}_j \rho^{\mathsf{XY}}
\Pi^\mathsf{X}_j/p_j$.

Such a generalization is needed because quantum information differs from
classical information in that the information that can be obtained from a
quantum system depends not only on its state but also on the choice of
measurements that is performed on it. So, in generalizing
$\mathbf{H}(\mathsf{Y}|\mathsf{X})$, we first had to choose a particular set of
one-dimensional orthogonal projectors $\big\{\Pi^\mathsf{X}_j\big\}$ acting on
the system $\mathsf{X}$.

It turns out that for general bipartite quantum states, the mutual information
is not identical to $\mathbf{J}(\mathsf{Y}:\mathsf{X})$ defined using
Eq.~(\ref{gen1}). The difference between $\mathbf{I}$ and $\mathbf{J}$ is called
\emph{quantum discord} and it is taken as a measure of non-classical
correlations in a quantum state \cite{PhysRevLett.88.017901}. 

\begin{figure}[htb]
\resizebox{8 cm}{5 cm}{\includegraphics{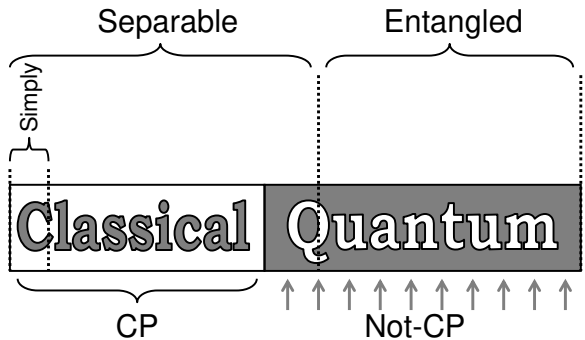}}
\caption{Quantum states of bipartite systems can be divided into two classes
based on their discord. States with quantum correlations have non-zero discord
while classically correlated states have zero discord. Separable states can have
quantum correlations while simply separable states have only classical
correlations. Not all quantum correlations are equivalent to entanglement. Also
shown is the nature of the dynamical maps induced by the unitary evolution of
the state of a system and its environment when the initial state belongs to each
class. Classically correlated states always lead to completely positive (CP)
maps while there are examples, indicated by the arrows, showing that states with
quantum correlations lead to not completely positive maps (Not-CP).}
\label{fig1}
\end{figure}

A quantum state with only classical correlations satisfies the condition $\rho^{\mathsf{XY}}= \sum_j \Pi^\mathsf{X}_j \rho^{\mathsf{XY}} \Pi^\mathsf{X}_j$. States of this form are a subset of the set of all separable states and the
subset includes all simply separable states. On the other hand, not all separable
states have only classical correlations. This implies that quantum correlations
must be taken to mean more than just entanglement. The information theoretic
characterization of quantum states based on the nature of the correlations
present is compared with the traditional division into separable and
entangled states in Fig.~\ref{fig1}.

In experiments, quantum systems are often initialized in desired states by
first performing a complete set of orthogonal projective measurements $\big\{
 \Pi_j \big\}$ on the system and then super-selecting the desired state from the
post-measurement state. After the measurements, the initial state of the system
and its environment has the form
\begin{equation} 
\label{initial}
\rho^{\mathcal{SE}} = \sum_j \Pi_j \rho^{\mathcal{SE}} \Pi_j =\sum_{j}p_j 
\Pi_j\otimes\tau_j, 
\end{equation}
where $\tau_j$ are density matrices for ${\mathcal{E}}$, $\big\{\Pi_j\big\}$ are a
complete set of orthogonal projectors on ${\mathcal{S}}$, $p_j\geq 0$ and 
$\sum_j p_j =1$. 

We now show that initial states of the system and the environment with only
classical correlations with respect to the system will always lead to completely positive
maps on the system under \emph{any} unitary evolution. Previously, only
simply separable states were known to lead to completely positivity reduced
dynamics for any choice of unitary evolution for the
combined state \cite{terno05,ziman06}.

We start from the classically correlated state from Eq.~(\ref{initial}). The initial state of the system is $\eta = \sum_j p_j \Pi_j$. 
 From Eq.~(\ref{TotalTrace}) we have
\begin{eqnarray*} 
\eta'_{rs}&=& \big[\mathfrak{B}\big]_{rr';ss'}\eta_{r's'} \\
&=& \mbox{Tr}_\mathcal{E} \bigg\{ \left[U\right]_{ra;r'a'} \Big(\sum_j p_j
\left[\Pi_j\right]_{r's'} [\tau_j]_{a'b'} \Big)[{U]^{*}_{sb;s'b'}}\bigg\}.
\end{eqnarray*}
Take the trace with respect to the environment by contracting indices $a$ and
$b$, 
\begin{equation*} 
\eta'_{rs}=\sum_j p_j \big[D^{kl}_{j}\big]_{rr'}  [\Pi_j]_{r's}
{\big[D^{kl}_{j}\big]^{*}_{ss'}},
\end{equation*}
where $[D^{kl}_{j}]_{rr'} \equiv [U]_{rl;r'a'}[\sqrt{\tau_{j}}]_{a'k}$. We have used
the fact that $\{\tau_j\}$ are positive to take their square root. After
combining indices $k$ and $l$ into a single index $\alpha$ we obtain, 
\begin{equation*} 
\eta'= \mathfrak{B}(\eta)=\sum_{j,\alpha} p_j D_{j}^{(\alpha)}  \Pi_j
{D_{j}^{(\alpha)}}^{\dagger}.
\end{equation*}
Expanding $D_{j}^{(\alpha)}$ as $\sum_m D_{m}^{(\alpha)}\delta_{jm}$ 
and using $\Pi^{2}_{j} = \Pi_j$ we obtain
\begin{equation*} 
\eta' =\sum_{j,\alpha} p_{j}  \Big(\sum_{m}
D_{m}^{(\alpha)}\delta_{jm} \Pi_{j}\Big) \Pi_{j} \Big(\sum_{n} \Pi_{j} 
\delta_{jn} {D_{n}^{(\alpha)}}^{\dagger} \Big). 
\end{equation*}
Now we can use the orthogonality of projectors, $\Pi_m \Pi_j
= \delta_{mj}\Pi_j$ to drop the dependency of $D_{j}^{(\alpha)}$ on index $j$
and write
\begin{equation*} 
\eta'=\sum_{j,\alpha} p_{j} \Big(\sum_{m} D_{m}^{(\alpha)}  \Pi_{m}\Big)
\Pi_{j} \Pi_{j}  \Pi_{j} \Big(\sum_{n}  {D_{n}^{(\alpha)}}\Pi_{n}
\Big)^{\dagger}. 
\end{equation*}
We can redefine $C^{(\alpha)} \equiv\sum_m D_{m}^{(\alpha)}\Pi_{m} $ to obtain,
\begin{equation} 
\label{final}
\eta^\prime=  \sum_\alpha  C^{(\alpha)} \Big(\sum_{j} p_{j}
\Pi_{j}\Big) {C^{(\alpha)}}^{\dagger}= \sum_\alpha C^{(\alpha)} \eta
{C^{(\alpha)}}^{\dagger}.
\end{equation}

Eq.~(\ref{final}) is identical to Eq.~(\ref{oper2}) showing that $\mathfrak{B}$
indeed is a completely positive map. This demonstrates that \emph{any} reduced unitary evolution of an
open system that is initially \emph{classically correlated} with its environment will
be given by a completely positive maps. The evolution of an open system
that has initial quantum correlations with the environment, on the other hand,
might lead to not completely positive maps as shown in Fig.~\ref{fig1}. 

Note that by specifying the initial state $\rho^{\mathcal{SE}}$ in
Eq.~(\ref{initial}) we have restricted to a subset of all possible initial system states. This subset is spanned by
the projectors $\big\{ \Pi_j \big\}$. The map $\mathfrak{B}$ from Eq.~(\ref{final}),
on the other hand, can be applied to any state of the system. Since the map is
completely positive it will map any system state to another valid state.
We will not, however, be able to understand the action of the map on states
outside the subset spanned by $\big\{ \Pi_j \big\}$ as coming from the
contraction of unitary evolution of the combined state in Eq.~(\ref{initial}).

Experimentally reconstructing dynamical maps corresponding to open quantum
evolution is called quantum process tomography \cite{chuang97a,Nielsen00a}. A
number of known initial states, sufficient to span the space of density matrices
of the system, are allowed to evolve as a result of an unknown process. The final
state corresponding to each initial state is then determined by quantum state
tomography. With the knowledge of the initial and corresponding final states the
linear dynamical map describing this unknown process is determined. The
complete set of initial states for the system is typically generated by 
creating a fiducial state and then applying controlled evolution on it to obtain
the other states. The fiducial state, in turn, is obtained by doing a
complete set of orthogonal measurements on the system as described earlier. 

We look at a representative quantum process tomography experiment on a
solid-state qubit performed by Howard et al. \cite{Howard05, Howard06} in the light of
the results presented above. In this experiment, the system
of interest is a qubit formed in a nitrogen vacancy defect in a diamond
lattice. Under ideal conditions, the experiment requires the initial state
of the qubit to be the pure state $|\phi \rangle \langle \phi|$, but in
reality the qubit is initialized in the state $\eta_{0}$
 with $p_0=\mbox{Tr}\left[|\phi\rangle\langle\phi|\eta_0 \right] =0.7$.

It is argued in \cite{Howard06} that the population considered was high enough
to effectively treat the initial state as  $|\phi\rangle\langle\phi|$. From this
state a complete set of linearly independent states is constructed
stochastically. This provides the set of initial states necessary to perform process
tomography on the decoherence occurring to the qubits. The map corresponding to
the decoherence process was found and it had negative eigenvalues, making it
not completely positive. The experimentally obtained not completely positive
maps were then discarded in favor of their ``closest" completely positive
counterparts \cite{havel03}. The occurrence of negative eigenvalues for the
dynamical maps was attributed to experimental errors. 

However, if we do not discard the negative eigenvalues as unphysical, this will
yield information about the initial preparation of the system.
If the system is indeed in a pure state $\eta_{0}\rightarrow
|\phi\rangle\langle\phi|$ initially then the combined state of the system
and its environment would necessarily be of the form $\rho^{\mathcal{SE}} =
|\phi\rangle\langle\phi|\otimes\tau$. Maps coming from such initially simply
separable states should be completely positive. This contradicts
what was found in the experiment. In addition to ruling out an initial simply separable state, we can now also
rule out initial states of the form
\begin{equation*} 
\rho^{\mathcal{SE}}=p_0|\phi\rangle\langle\phi|\otimes\tau^\prime+\left(1-p_0\right)|\phi_\bot\rangle\langle\phi_\bot|\otimes\tau^{\prime\prime},
\end{equation*}
with $\langle \phi |\phi_\bot\rangle=0$, even though states of this form are
consistent with the measured population $p_0$. However, a state like this only
has classical correlations, and we know that the map induced by any unitary
evolution of such a state should be completely positive.

The not completely positive map found in this experiment could be interpreted as
an indication that the initial state of the system is not just classically
correlated with the environment. Given that the qubit is in a large crystal
lattice, it is perhaps not very surprising that it had quantum correlations with
the surrounding environment.

We propose that if after performing quantum process tomography a not completely
positive map is found, this should be considered as a signature that the system
had quantum correlations with the environment. Our definition of quantum
correlation is different from the ones considered in other previous studies by
other authors \cite{terno05,ziman06}.

In conclusion, we have studied the effect of initial correlations with the
environment on the complete positivity of dynamical maps that describe the
open-systems evolution. We proved that if there are only classical correlations
in the state of the system and its environment, as indicated by zero discord,
then the maps induced by any unitary evolution of the combined state must be
completely positive for any unitary transformation. This result is more general
than the previously known result for simply separable initial states. 


{\bf Acknowledgments}: The authors wish to thank T. F. Jordan and W. Zurek for useful
comments and discussions. Anil Shaji acknowledges the support of the US Office
of Naval Research through Contract No.~N00014-03-1-0426.

\bibliography{discordCPmapsPRL16}

\begin{thebibliography}{32}
\expandafter\ifx\csname natexlab\endcsname\relax\def\natexlab#1{#1}\fi
\expandafter\ifx\csname bibnamefont\endcsname\relax
  \def\bibnamefont#1{#1}\fi
\expandafter\ifx\csname bibfnamefont\endcsname\relax
  \def\bibfnamefont#1{#1}\fi
\expandafter\ifx\csname citenamefont\endcsname\relax
  \def\citenamefont#1{#1}\fi
\expandafter\ifx\csname url\endcsname\relax
  \def\url#1{\texttt{#1}}\fi
\expandafter\ifx\csname urlprefix\endcsname\relax\def\urlprefix{URL }\fi
\providecommand{\bibinfo}[2]{#2}
\providecommand{\eprint}[2][]{\url{#2}}

\bibitem[{\citenamefont{Davies}(1976)}]{davies76a}
\bibinfo{author}{\bibfnamefont{E.~B.} \bibnamefont{Davies}},
  \emph{\bibinfo{title}{Quantum theory of open systems}}
  (\bibinfo{publisher}{Academic Press}, \bibinfo{address}{New York},
  \bibinfo{year}{1976}).

\bibitem[{\citenamefont{Spohn}(1980)}]{spohn80a}
\bibinfo{author}{\bibfnamefont{H.}~\bibnamefont{Spohn}}, \bibinfo{journal}{Rev.
  Mod. Phys.} \textbf{\bibinfo{volume}{53}}, \bibinfo{pages}{569}
  (\bibinfo{year}{1980}).

\bibitem[{\citenamefont{Breuer and Petruccione}(2002)}]{breuer02a}
\bibinfo{author}{\bibfnamefont{H.~P.} \bibnamefont{Breuer}} \bibnamefont{and}
  \bibinfo{author}{\bibfnamefont{F.}~\bibnamefont{Petruccione}},
  \emph{\bibinfo{title}{The theory of open quantum systems}}
  (\bibinfo{publisher}{Oxford university press}, \bibinfo{address}{New York},
  \bibinfo{year}{2002}).

\bibitem[{\citenamefont{Sudarshan et~al.}(1961)\citenamefont{Sudarshan,
  Mathews, and Rau}}]{PhysRev.121.920}
\bibinfo{author}{\bibfnamefont{E.~C.~G.} \bibnamefont{Sudarshan}},
  \bibinfo{author}{\bibfnamefont{P.~M.} \bibnamefont{Mathews}},
  \bibnamefont{and} \bibinfo{author}{\bibfnamefont{J.}~\bibnamefont{Rau}},
  \bibinfo{journal}{Phys. Rev.} \textbf{\bibinfo{volume}{121}},
  \bibinfo{pages}{920} (\bibinfo{year}{1961}).

\bibitem[{\citenamefont{Choi}(1972)}]{choi72a}
\bibinfo{author}{\bibfnamefont{M.~D.} \bibnamefont{Choi}},
  \bibinfo{journal}{Can. J. Math.} \textbf{\bibinfo{volume}{24}},
  \bibinfo{pages}{520} (\bibinfo{year}{1972}).

\bibitem[{\citenamefont{Choi}(1975)}]{choi75}
\bibinfo{author}{\bibfnamefont{M.~D.} \bibnamefont{Choi}},
  \bibinfo{journal}{Linear Algebra and Appl.} \textbf{\bibinfo{volume}{10}},
  \bibinfo{pages}{285} (\bibinfo{year}{1975}).

\bibitem[{\citenamefont{Kraus}(1983)}]{kraus83b}
\bibinfo{author}{\bibfnamefont{K.}~\bibnamefont{Kraus}},
  \emph{\bibinfo{title}{States, Effects and Operations: Fundamental Notions of
  Quantum Theory}}, vol. \bibinfo{volume}{190} of
  \emph{\bibinfo{series}{Lecture notes in Physics}}
  (\bibinfo{publisher}{Spring-Verlag}, \bibinfo{address}{New York},
  \bibinfo{year}{1983}).

\bibitem[{\citenamefont{Sudarshan}(1986)}]{sudarshan86}
\bibinfo{author}{\bibfnamefont{E.~C.~G.} \bibnamefont{Sudarshan}},
  \emph{\bibinfo{title}{Quantum Measurement and Dynamical Maps in `From $SU(3)$
  to Gravity'}} (\bibinfo{publisher}{Cambridge University Press},
  \bibinfo{year}{1986}).

\bibitem[{\citenamefont{Sudarshan}(2003)}]{SudChaos}
\bibinfo{author}{\bibfnamefont{E.~C.~G.} \bibnamefont{Sudarshan}},
  \bibinfo{journal}{Chaos, Solitons Fractals} \textbf{\bibinfo{volume}{16}},
  \bibinfo{pages}{369} (\bibinfo{year}{2003}).

\bibitem[{\citenamefont{Nielsen and Chuang}(200)}]{Nielsen00a}
\bibinfo{author}{\bibfnamefont{M.~A.} \bibnamefont{Nielsen}} \bibnamefont{and}
  \bibinfo{author}{\bibfnamefont{I.~L.} \bibnamefont{Chuang}},
  \emph{\bibinfo{title}{Quantum Computation and Quantum Information}}
  (\bibinfo{publisher}{Cambridge University Press},
  \bibinfo{address}{Cambridge, UK}, \bibinfo{year}{200}).

\bibitem[{\citenamefont{Pechukas}(1994)}]{pechukas94a}
\bibinfo{author}{\bibfnamefont{P.}~\bibnamefont{Pechukas}},
  \bibinfo{journal}{Phys. Rev. Lett.} \textbf{\bibinfo{volume}{73}},
  \bibinfo{pages}{1060} (\bibinfo{year}{1994}).

\bibitem[{\citenamefont{Jordan et~al.}(2005)\citenamefont{Jordan, Shaji, and
  Sudarshan}}]{jordan05a}
\bibinfo{author}{\bibfnamefont{T.~F.} \bibnamefont{Jordan}},
  \bibinfo{author}{\bibfnamefont{A.}~\bibnamefont{Shaji}}, \bibnamefont{and}
  \bibinfo{author}{\bibfnamefont{E.~C.~G.} \bibnamefont{Sudarshan}},
  \bibinfo{journal}{Phys. Rev. A.} \textbf{\bibinfo{volume}{73}},
  \bibinfo{pages}{32104} (\bibinfo{year}{2005}).

\bibitem[{\citenamefont{Shaji}(2005)}]{shaji05dis}
\bibinfo{author}{\bibfnamefont{A.}~\bibnamefont{Shaji}}, Ph.D. thesis,
  \bibinfo{school}{The University of Texas at Austin} (\bibinfo{year}{2005}).

\bibitem[{\citenamefont{Jordan et~al.}(2004)\citenamefont{Jordan, Shaji, and
  Sudarshan}}]{jordan:052110}
\bibinfo{author}{\bibfnamefont{T.~F.} \bibnamefont{Jordan}},
  \bibinfo{author}{\bibfnamefont{A.}~\bibnamefont{Shaji}}, \bibnamefont{and}
  \bibinfo{author}{\bibfnamefont{E.~C.~G.} \bibnamefont{Sudarshan}},
  \bibinfo{journal}{Phys. Rev. A} \textbf{\bibinfo{volume}{70}},
  \bibinfo{eid}{052110} (\bibinfo{year}{2004}).

\bibitem[{\citenamefont{Carteret et~al.}(2005)\citenamefont{Carteret, Terno,
  and Zyczkowski}}]{terno05}
\bibinfo{author}{\bibfnamefont{H.}~\bibnamefont{Carteret}},
  \bibinfo{author}{\bibfnamefont{D.}~\bibnamefont{Terno}}, \bibnamefont{and}
  \bibinfo{author}{\bibfnamefont{K.}~\bibnamefont{Zyczkowski}},
  \bibinfo{journal}{arXiv:quant-ph} \bibinfo{eid}{0512167}
  (\bibinfo{year}{2005}).

\bibitem[{\citenamefont{Shabani and Lidar}(2006)}]{shabani06a}
\bibinfo{author}{\bibfnamefont{A.}~\bibnamefont{Shabani}} \bibnamefont{and}
  \bibinfo{author}{\bibfnamefont{D.~A.} \bibnamefont{Lidar}},
  \bibinfo{journal}{arXiv:quant-ph} \bibinfo{eid}{0610028}
  (\bibinfo{year}{2006}).

\bibitem[{\citenamefont{Ziman}(2006)}]{ziman06}
\bibinfo{author}{\bibfnamefont{M.}~\bibnamefont{Ziman}},
  \bibinfo{journal}{arXiv:quant-ph} \bibinfo{eid}{0603166}
  (\bibinfo{year}{2006}).

\bibitem[{\citenamefont{Ollivier and Zurek}(2001)}]{PhysRevLett.88.017901}
\bibinfo{author}{\bibfnamefont{H.}~\bibnamefont{Ollivier}} \bibnamefont{and}
  \bibinfo{author}{\bibfnamefont{W.~H.} \bibnamefont{Zurek}},
  \bibinfo{journal}{Phys. Rev. Lett.} \textbf{\bibinfo{volume}{88}},
  \bibinfo{pages}{017901} (\bibinfo{year}{2001}).

\bibitem[{\citenamefont{Henderson and Vedral}(2001)}]{henderson01a}
\bibinfo{author}{\bibfnamefont{L.}~\bibnamefont{Henderson}} \bibnamefont{and}
  \bibinfo{author}{\bibfnamefont{V.}~\bibnamefont{Vedral}},
  \bibinfo{journal}{J. Phys. A} \textbf{\bibinfo{volume}{34}},
  \bibinfo{pages}{6899} (\bibinfo{year}{2001}).

\bibitem[{\citenamefont{Peres}(1996)}]{peres96a}
\bibinfo{author}{\bibfnamefont{A.}~\bibnamefont{Peres}},
  \bibinfo{journal}{Phys. Rev. Lett.} \textbf{\bibinfo{volume}{77}},
  \bibinfo{pages}{1413} (\bibinfo{year}{1996}).

\bibitem[{\citenamefont{Rodriguez et~al.}(2006)\citenamefont{Rodriguez, Shaji,
  and Sudarshan}}]{rodr}
\bibinfo{author}{\bibfnamefont{C.}~\bibnamefont{Rodriguez}},
  \bibinfo{author}{\bibfnamefont{A.}~\bibnamefont{Shaji}}, \bibnamefont{and}
  \bibinfo{author}{\bibfnamefont{E.~C.~G.} \bibnamefont{Sudarshan}},
  \bibinfo{journal}{arXiv:quant-ph} \bibinfo{eid}{0504051}
  (\bibinfo{year}{2006}).

\bibitem[{\citenamefont{Jordan et~al.}(2006)\citenamefont{Jordan, Shaji, and
  Sudarshan}}]{jordan06a}
\bibinfo{author}{\bibfnamefont{T.~F.} \bibnamefont{Jordan}},
  \bibinfo{author}{\bibfnamefont{A.}~\bibnamefont{Shaji}}, \bibnamefont{and}
  \bibinfo{author}{\bibfnamefont{E.~C.~G.} \bibnamefont{Sudarshan}},
  \bibinfo{journal}{Phys. Rev. A.} \textbf{\bibinfo{volume}{73}},
  \bibinfo{pages}{12106} (\bibinfo{year}{2006}).

\bibitem[{\citenamefont{Jordan}(2005)}]{jordan:034101}
\bibinfo{author}{\bibfnamefont{T.~F.} \bibnamefont{Jordan}},
  \bibinfo{journal}{Phys. Rev. A} \textbf{\bibinfo{volume}{71}},
  \bibinfo{eid}{034101} (\bibinfo{year}{2005}).

\bibitem[{\citenamefont{Werner}(1989)}]{PhysRevA.40.4277}
\bibinfo{author}{\bibfnamefont{R.~F.} \bibnamefont{Werner}},
  \bibinfo{journal}{Phys. Rev. A} \textbf{\bibinfo{volume}{40}},
  \bibinfo{pages}{4277} (\bibinfo{year}{1989}).

\bibitem[{\citenamefont{Bennett et~al.}(1999)\citenamefont{Bennett, Shor,
  Smolin, and Thapliyal}}]{bennett99b}
\bibinfo{author}{\bibfnamefont{C.~H.} \bibnamefont{Bennett}},
  \bibinfo{author}{\bibfnamefont{P.~W.} \bibnamefont{Shor}},
  \bibinfo{author}{\bibfnamefont{J.~A.} \bibnamefont{Smolin}},
  \bibnamefont{and} \bibinfo{author}{\bibfnamefont{A.~V.}
  \bibnamefont{Thapliyal}}, \bibinfo{journal}{Phys. Rev. Lett.}
  \textbf{\bibinfo{volume}{83}}, \bibinfo{pages}{3081} (\bibinfo{year}{1999}).

\bibitem[{\citenamefont{Bennett et~al.}(2002)\citenamefont{Bennett, Shor,
  Smolin, and Thapliyal}}]{bennett02a}
\bibinfo{author}{\bibfnamefont{C.~H.} \bibnamefont{Bennett}},
  \bibinfo{author}{\bibfnamefont{P.~W.} \bibnamefont{Shor}},
  \bibinfo{author}{\bibfnamefont{J.~A.} \bibnamefont{Smolin}},
  \bibnamefont{and} \bibinfo{author}{\bibfnamefont{A.~V.}
  \bibnamefont{Thapliyal}}, \bibinfo{journal}{IEEE Trans. Information Theory}
  \textbf{\bibinfo{volume}{48}}, \bibinfo{pages}{2637} (\bibinfo{year}{2002}).

\bibitem[{\citenamefont{Shannon}(1948)}]{shannon48a}
\bibinfo{author}{\bibfnamefont{C.~E.} \bibnamefont{Shannon}},
  \bibinfo{journal}{Bell System Tech. J.} \textbf{\bibinfo{volume}{27}},
  \bibinfo{pages}{379} (\bibinfo{year}{1948}).

\bibitem[{\citenamefont{von Neumann}(1955)}]{vonNeu}
\bibinfo{author}{\bibfnamefont{J.}~\bibnamefont{von Neumann}},
  \emph{\bibinfo{title}{Mathematical Foundations of Quantum Mechanics}}
  (\bibinfo{publisher}{Princeton University}, \bibinfo{address}{Princeton},
  \bibinfo{year}{1955}).

\bibitem[{\citenamefont{Chuang and Nielsen}(1997)}]{chuang97a}
\bibinfo{author}{\bibfnamefont{I.~L.} \bibnamefont{Chuang}} \bibnamefont{and}
  \bibinfo{author}{\bibfnamefont{M.~A.} \bibnamefont{Nielsen}},
  \bibinfo{journal}{J. Mod. Optics} \textbf{\bibinfo{volume}{44}},
  \bibinfo{pages}{2455} (\bibinfo{year}{1997}).

\bibitem[{\citenamefont{Howard et~al.}(2005)\citenamefont{Howard, Twamley,
  Wittmann, Gaebel, Jelezko, and Wrachtrup}}]{Howard05}
\bibinfo{author}{\bibfnamefont{M.}~\bibnamefont{Howard}},
  \bibinfo{author}{\bibfnamefont{J.}~\bibnamefont{Twamley}},
  \bibinfo{author}{\bibfnamefont{C.}~\bibnamefont{Wittmann}},
  \bibinfo{author}{\bibfnamefont{T.}~\bibnamefont{Gaebel}},
  \bibinfo{author}{\bibfnamefont{F.}~\bibnamefont{Jelezko}}, \bibnamefont{and}
  \bibinfo{author}{\bibfnamefont{J.}~\bibnamefont{Wrachtrup}},
  \bibinfo{journal}{arXiv:quant-ph} \bibinfo{eid}{0503153}
  (\bibinfo{year}{2005}).

\bibitem[{\citenamefont{Howard et~al.}(2006)\citenamefont{Howard, Twamley,
  Wittmann, Gaebel, Jelezko, and Wrachtrup}}]{Howard06}
\bibinfo{author}{\bibfnamefont{M.}~\bibnamefont{Howard}},
  \bibinfo{author}{\bibfnamefont{J.}~\bibnamefont{Twamley}},
  \bibinfo{author}{\bibfnamefont{C.}~\bibnamefont{Wittmann}},
  \bibinfo{author}{\bibfnamefont{T.}~\bibnamefont{Gaebel}},
  \bibinfo{author}{\bibfnamefont{F.}~\bibnamefont{Jelezko}}, \bibnamefont{and}
  \bibinfo{author}{\bibfnamefont{J.}~\bibnamefont{Wrachtrup}},
  \bibinfo{journal}{New J. Phys.} \textbf{\bibinfo{volume}{8}},
  \bibinfo{pages}{33} (\bibinfo{year}{2006}).

\bibitem[{\citenamefont{Havel}(2003)}]{havel03}
\bibinfo{author}{\bibfnamefont{T.~F.} \bibnamefont{Havel}},
  \bibinfo{journal}{J. Math. Phys.} \textbf{\bibinfo{volume}{44}},
  \bibinfo{pages}{534} (\bibinfo{year}{2003}).

\end{thebibliography}

\end{document}